\documentclass[prd,aps,preprint,tightenlines,superscriptaddress,nofootinbib,showpacs,showkeys]{revtex4}
\usepackage{float}
\usepackage{mathrsfs}
\usepackage{amsfonts}
\usepackage{array}
\usepackage{epsfig}
\usepackage{epstopdf}
\usepackage{amsmath}    
\usepackage{amssymb}   
\usepackage{graphicx}   
\usepackage{verbatim}   
\usepackage{color}      
\usepackage{subfigure}  
\usepackage{hyperref}   
\usepackage{url}
\usepackage{CJK}
\usepackage{array,multirow}

\usepackage{tikz}
\usetikzlibrary{trees}
\usetikzlibrary{decorations.pathmorphing}
\usetikzlibrary{decorations.markings}

  \definecolor{jblue}  {RGB}{20,50,100}
  \definecolor{npurple}  {RGB} {153, 51, 204}
  \definecolor{wred}   {RGB}{217,0,56}
  \definecolor{white}   {RGB}{255,255,255}

  \definecolor{korange}   {RGB}{235, 80,  43}
  \definecolor{korange2}   {RGB}{245, 100,  63}
  \definecolor{kyelloworange}   {RGB}{255, 210,  110}
  \definecolor{kyelloworange2}   {RGB}{240, 170,  90}
  \definecolor{kred}   {RGB}{204,  102, 153}
  \definecolor{kpurple}   {RGB}{153,  61, 190}
  \definecolor{kpurplelight}   {RGB}{213,  161, 230}

	\tikzset{
	  photon/.style={decorate, decoration={snake}, draw=npurple,very thick},
	  boson/.style={decorate, decoration={snake}, draw=npurple,very thick},
	  electron/.style={draw=jblue,very thick, postaction={decorate},
	           decoration={markings,mark=at position .55 with {\arrow[draw=jblue]{>}}}
	  },
	  electron2/.style={draw=jblue,very thick, postaction={decorate},
	           decoration={markings,mark=at position .55 with {\arrow[draw=jblue]{<}}}
	  },
	  fermion/.style={draw=jblue,very thick, postaction={decorate},
	            decoration={markings,mark=at position .55 with {\arrow[draw=jblue]{}}}
	  },
	  gluon/.style={decorate, draw=korange,very thick, 
	    decoration={coil,amplitude=4pt, segment length=6pt}},
	  higgs/.style={draw=wred,very thick, postaction={decorate},
	           decoration={markings,mark=at position .55 with {\arrow[draw=wred]{}}}
	  },
	  nothing/.style={draw=white,very thick}
	}

\def\gev{{\ifmmode{\mathrm{\:GeV}}\else${\mathrm{\:GeV}}$\fi}}
\def\tev{{\ifmmode{\mathrm{\:TeV}}\else${\mathrm{\:TeV}}$\fi}}

\begin{document}
\begin{CJK*}{UTF8}{gbsn}
	
\title{Next-to-leading order QCD  and  electroweak  corrections to Higgs-strahlung processes at the LHC }

\author{Pazilet Obul}
\email{pazilet.obul@hotmail.com}
\affiliation{
School of Physics Science and Technology, Xinjiang University,
 Urumqi, Xinjiang 830046 China }
\affiliation{
Center for Theoretical Physics, Xinjiang University,\\
 Urumqi, Xinjiang 830046 China }
\author{Sayipjamal Dulat}
\email{sdulat@hotmail.com}
\affiliation{
School of Physics Science and Technology, Xinjiang University,
 Urumqi, Xinjiang 830046 China }
\affiliation{
Center for Theoretical Physics, Xinjiang University,\\
 Urumqi, Xinjiang 830046 China }
\author{Tie-Jiun Hou }
\affiliation{
School of Physics Science and Technology, Xinjiang University,
 Urumqi, Xinjiang 830046 China }
\affiliation{
Center for Theoretical Physics, Xinjiang University,\\
 Urumqi, Xinjiang 830046 China }
\author{Ablikim Tursun}
\affiliation{
School of Physics Science and Technology, Xinjiang University,
 Urumqi, Xinjiang 830046 China }
\affiliation{
Center for Theoretical Physics, Xinjiang University,\\
 Urumqi, Xinjiang 830046 China }
\author{Nijat Yalkun}
\affiliation{
School of Physics Science and Technology, Xinjiang University,
 Urumqi, Xinjiang 830046 China }
\affiliation{
Center for Theoretical Physics, Xinjiang University,\\
 Urumqi, Xinjiang 830046 China }

\begin{abstract}

In this paper we calculate the total and fiducial cross sections as well
as differential distributions for the Higgs-strahlung  or VH process $p p \to VH \to  l\nu_l/l^- l^+ + H$, (V = W or Z, l=e,$\mu$)  including QCD and electro-weak  corrections up to next-to-leading order  before and after
reweighting photon PDFs of NNPDF2.3qed, NNPDF3.0qed, MRST2004qed, CT14QEDinc,  and LUXqed at the LHC with 13 TeV and Higgs boson mass $\ M_{H}=125$ GeV.
The predictions from the various photon PDFs  before and after reweighting against each other are in good agreement. The photon PDF uncertainties of the photon-induced cross sections decrease significantly with the reweighting PDFs.

\end{abstract}


\maketitle
\tableofcontents

\newpage
\section{Introduction} \label{intro}

The Standard Model (SM) predicts  the Higgs boson, that
is, the remnant of the electroweak (EW) symmetry-breaking mechanism that generates the
gauge boson and fermion masses \cite{review}.
One of the most important Higgs boson production mechanisms at hadron colliders is
the Higgs-strahlung process, i.e. the associated production of Higgs bosons and weak gauge bosons,
\begin{equation}
pp\to W H+X  \to H \, \nu_l l +X   \quad\mbox{and}\quad pp\to ZH+X \to H \,l^+ l^-  +X
\label{eq:procs}
\end{equation}
In order to determine the Higgs boson properties with high precision
at the LHC, it is necessary to calculate the higher order quantum chromodynamics (QCD) and EW corrections. One particular EW correction of interest is that due to photons coming from the proton in the initial state. It is thus  necessary to use photon parton distribution functions (PDFs) both for consistency when EW corrections are included, and because photon-induced processes can become important at high energies. So far, a number of PDF groups, namely, the MRST, NNPDF and CTEQ collaborations, have introduced photon PDFs along
with PDF evolution at leading order (LO) in QED and next-to-leading order (NLO) or next-to-next-to-leading order (NNLO) in QCD.
 Reference~\cite{:1991.han} calculated NLO QCD corrections to the $pp \to W^\pm H + X$ (sum of $W^+H$ and $W^-H$) and $pp \to ZH + X$ total cross sections at the  LHC at  16~TeV and 40 TeV, and discussed in detail the total cross section dependence on the choice of factorization scale and for three different sets of PDFs (MT, KMRS and DFLM) within the DIS scheme and $\overline{MS}$ scheme for $m_H = 100$ GeV.
References~\cite{arXiv:0306234,arXiv:0402003} presented the impact of the EW corrections on the cross
section predictions for the processes $pp/p\bar p \to W^\pm H + X$  and $pp/p\bar p \to ZH + X$ at the Tevatron $(\sqrt s = 1.96~TeV)$ and LHC $(\sqrt s = 14~TeV)$ for the three different input parameter schemes ($G_\mu$, $\alpha(M_Z)$, and $\alpha(0)$-schemes), and studied the theoretical uncertainties induced by factorization and renormalization scale dependence and by the PDFs by using the CTEQ6L1 and CTEQ6M \cite{CTEQ6} PDFs, respectively.
The EW corrections in the $G_\mu$ and  $\alpha(M_Z)$-schemes are significant and reduce the cross section by
5-9\% and by 10-15\%, respectively. The EW corrections in the $\alpha(0)$-scheme are rather small.
 References~\cite{arXiv:0307206,arXiv:0402003} discussed the NNLO QCD, i.e. the $O(\alpha^2_s)$, corrections to the $pp/p\bar p  \to W^\pm H $ and $pp/p\bar p  \to ZH$ production cross sections at the Tevatron at 1.96~TeV and  LHC at 14~TeV for Higgs boson masses $M_{H} \lesssim 200-300$ GeV using the MRST~\cite{MRST} PDFs. For $M_H = 120 GeV$,  the renormalization and factorization scale uncertainty at the LHC at 14~TeV drops from 10\% at LO to 5\% at NLO, and to 2\% at NNLO. At the Tevatron and for
the same Higgs boson mass, the scale uncertainty drops from 20\% at LO to 7\% at NLO, and to 3\% at NNLO.
A. Denner et al~\cite{arXiv:1112.5142}  have discussed the impact of EW radiative corrections to the Higgs-strahlung
 processes at the Tevatron (1.96~TeV) and LHC (7~TeV and 14~TeV) within the $G_\mu$-scheme  for $M_H = 120 GeV$ using the HAWK Monte Carlo program.
The LHC Higgs Cross Section Working Group~\cite{2016-008cc} provided an update of the total and fiducial cross-section, together with theoretical uncertainties from renormalization and factorization scales,  of the process $pp \to W^\pm /Z + H  \to l\nu_l \//l^+l^- +H$ including NNLO QCD  and NLO EW corrections for different proton-proton collision energies $(\sqrt s = 7 TeV, 8 TeV, 13 TeV, 14 TeV)$ for a Higgs boson mass $M_H = 125 GeV$.
Reference~\cite{Ababekri:2016kkj} studied the impact of the CMS measurements of W boson pair production  at the LHC at 8~TeV on the NNPDF2.3qed, NNPDF3.0qed~\cite{Ball:2014uwa}, CT14QEDinc~\cite{Schmidt:2015zda}, and LUXqed~\cite{Manohar:2016nzj} photon PDFs, and found that the data strongly suggest that the NNPDF photon error band should be significantly reduced.
Therefore, it is necessary to  re-examine the physics effects induced by photon-initiated process.
In this paper, by using various PDFs, we update the predictions  for the processes $pp \to W^\pm /Z + H  \to l\nu_l \//l^+l^- +H$ at the  LHC at 13 TeV and Higgs boson mass $M_H = 125$ GeV, including the NLO QCD and EW corrections using the most recent PDFs.

The paper is organized as follows. In Section \ref{Results}, first we give the Feynman diagrams for the Higgs-strahlung processes $pp \to W^\pm /Z + H  \to l\nu_l \//l^+l^- +H$  at LO and NLO.
Then we provide our numerical results for these processes at the
LHC with center-of-mass energy 13 TeV using the Monte Carlo program HAWK~\cite{hawk} by adopting the NNPDF2.3qed, NNPDF3.0qed, MRST2004qed, CT14QEDinc, and NNLO LUXqed PDFs, as well as their reweighting photon PDFs.
We obtain reweighting photon PDFs by using the reweighting method \cite{Giele:1998gw,Ball:2010gb, Ball:2011gg,Sato:2013ika}  and  the LHC CMS 8 TeV data~\cite{Khachatryan:2016mud}  as in Ref.~\cite{Ababekri:2016kkj}. Our conclusion is given in Section \ref{conclusion}.

\section{Results}\label{Results}
\subsection{Feynman Diagrams}\label{diagrams}

\begin{figure}[!htb]
	\centering
	\begin{tikzpicture}[scale=0.90,
					thick,
			level/.style={level distance=3.15cm, line width=0.4mm},
			level 2/.style={sibling angle=60},
			level 3/.style={sibling angle=60},
			level 4/.style={level distance=1.4cm, sibling angle=60}
	]

	\draw[electron2] (0,0) -- (-1,1) ;
	\draw[electron] (0,0) -- (-1,-1) ;
	\draw[boson] (0,0) -- (1,0) ;
	\draw[boson] (1,0) -- (1.3,1) ;
	\draw[higgs,dashed] (1,0) -- (2,-1) ;
	\draw[electron2] (1.3,1) -- (2,0.3) ;
	\draw[electron] (1.3,1) -- (2,1.7) ;
        \node[draw=none,fill=none] at (0.6,-0.4){$W$}  ; 
        \node[draw=none,fill=none] at (2.2,-0.7){$H$}  ; 
        \node[draw=none,fill=none] at (0.8,0.7){$W$}  ; 
         \node[draw=none,fill=none] at (-1.2,0.7){$q_{i,u}$}  ; 
        \node[draw=none,fill=none] at (-1.2,-0.6){$\overline{q}_{j,d}$}  ; 
         \node[draw=none,fill=none] at (2.2,0.6){$l^{+}$}  ; 
        \node[draw=none,fill=none] at (2.2,1.5){$\nu_{l}$}  ; 
	
	\fill[red]   (0,0)  circle (0.05cm);
	\draw[korange]         (0,0)  circle (0.05cm);
	\fill[red]   (1,0)  circle (0.05cm);
	\draw[korange]         (1,0)  circle (0.05cm);
\fill[red]   (1.3,1)  circle (0.05cm);
	\draw[korange]         (1.3,1)  circle (0.05cm);
\end{tikzpicture}
\hspace{0.5cm}
\begin{tikzpicture}[scale=0.90,
					thick,
			level/.style={level distance=3.15cm, line width=0.4mm},
			level 2/.style={sibling angle=60},
			level 3/.style={sibling angle=60},
			level 4/.style={level distance=1.4cm, sibling angle=60}
	]

	\draw[electron2] (0,0) -- (-1,1) ;
	\draw[electron] (0,0) -- (-1,-1) ;
	\draw[boson] (0,0) -- (1,0) ;
	\draw[boson] (1,0) -- (1.3,1) ;
	\draw[higgs,dashed] (1,0) -- (2,-1) ;
	\draw[electron2] (1.3,1) -- (2,0.3) ;
	\draw[electron] (1.3,1) -- (2,1.7) ;
        \node[draw=none,fill=none] at (0.6,-0.4){$W$}  ; 
        \node[draw=none,fill=none] at (2.2,-0.7){$H$}  ; 
        \node[draw=none,fill=none] at (0.8,0.7){$W$}  ; 
         \node[draw=none,fill=none] at (-1.2,0.7){$q_{i,d}$}  ; 
        \node[draw=none,fill=none] at (-1.2,-0.6){$\overline{q}_{j,u}$}  ; 
         \node[draw=none,fill=none] at (2.2,0.6){$\bar{\nu}_{l}$}  ; 
        \node[draw=none,fill=none] at (2.2,1.4){$l^{-}$}  ; 
	
	\fill[red]   (0,0)  circle (0.05cm);
	\draw[korange]         (0,0)  circle (0.05cm);
	\fill[red]   (1,0)  circle (0.05cm);
	\draw[korange]         (1,0)  circle (0.05cm);
\fill[red]   (1.3,1)  circle (0.05cm);
	\draw[korange]         (1.3,1)  circle (0.05cm);
\end{tikzpicture}
\hspace{0.5cm}
\begin{tikzpicture}[scale=0.90,
					thick,
			level/.style={level distance=3.15cm, line width=0.4mm},
			level 2/.style={sibling angle=60},
			level 3/.style={sibling angle=60},
			level 4/.style={level distance=1.4cm, sibling angle=60}
	]

	\draw[electron2] (0,0) -- (-1,1) ;
	\draw[electron] (0,0) -- (-1,-1) ;
	\draw[boson] (0,0) -- (1,0) ;
	\draw[boson] (1,0) -- (1.3,1) ;
	\draw[higgs,dashed] (1,0) -- (2,-1) ;
	\draw[electron2] (1.3,1) -- (2,0.3) ;
	\draw[electron] (1.3,1) -- (2,1.7) ;
        \node[draw=none,fill=none] at (0.6,-0.4){$Z$}  ; 
        \node[draw=none,fill=none] at (2.2,-0.7){$H$}  ; 
        \node[draw=none,fill=none] at (0.8,0.7){$Z$}  ; 
         \node[draw=none,fill=none] at (-1,0.7){$q_{i}$}  ; 
        \node[draw=none,fill=none] at (-1,-0.6){$\overline{q}_{i}$}  ; 
         \node[draw=none,fill=none] at (2.2,0.6){$l^{+}$}  ; 
        \node[draw=none,fill=none] at (2.2,1.4){$l^{-}$}  ; 
	
	\fill[red]   (0,0)  circle (0.05cm);
	\draw[korange]         (0,0)  circle (0.05cm);
	\fill[red]   (1,0)  circle (0.05cm);
	\draw[korange]         (1,0)  circle (0.05cm);
\fill[red]   (1.3,1)  circle (0.05cm);
	\draw[korange]         (1.3,1)  circle (0.05cm);
\end{tikzpicture}
   	\caption{ Feynman diagram for Higgs-strahlung processes at LO. The straight, wavy, and dashed
lines denote fermions, weak gauge bosons, and Higgs boson, respectively.}
	\label{fig:feynmanHiggs}
\end{figure}
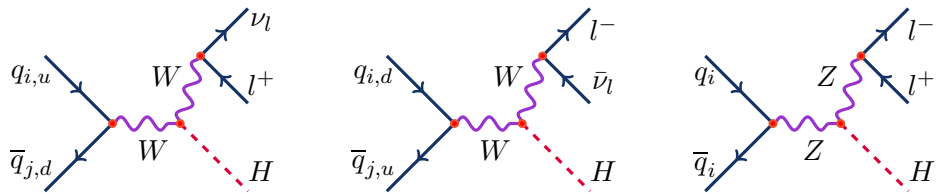

The LO Feynman diagrams for Higgs boson production in association with weak gauge
bosons $V=W^\pm,Z$, at the proton-antiproton collider Tevatron and  proton-proton collider LHC, are shown in Fig.~\ref{fig:feynmanHiggs}.
The corresponding partonic processes are
$q_{i,u} \, \, \bar q_{j,d} \,\to   H W^+  \to\, H l^+ \nu_l$ \; and
$q_{i,d} \, \, \bar q_{j,u} \,\to  H W^-  \to\, H l^- \bar\nu_l$, where $q_{i,u}$ and $q_{i,d}$ denote  up- and down-type quarks of the $i$th generation,
as well as $q_i \, \, \bar q_i \,\to  H Z  \to\, H l^+ l^- $, where $q_i$ denotes any light quark of the $i$th generation.

\begin{figure}[!htb]
	\centering
\begin{flushleft}
	Vertex diagrams:
	\end{flushleft}

\vspace{0.09cm}

\begin{tikzpicture}[scale=0.90,
					thick,
			level/.style={level distance=3.15cm, line width=0.4mm},
			level 2/.style={sibling angle=60},
			level 3/.style={sibling angle=60},
			level 4/.style={level distance=1.4cm, sibling angle=60}
	]

	\draw[electron2] (0,0) -- (-1,1) ;
	\draw[electron] (0,0) -- (-1,-1) ;
	\draw[boson] (-0.5,0.5) -- (-0.5,-0.5) ;
	\draw[boson] (0,0) -- (1,0) ;
	\draw[boson] (1,0) -- (2,1) ;
	\draw[higgs,dashed] (1,0) -- (2,-1) ;
	\node[draw=none,fill=none] at (-1,0){$V,\gamma$}  ; 
        \node[draw=none,fill=none] at (0.5,0.3){$V$}  ; 
          \node[draw=none,fill=none] at (2.1,0.6){$V$}  ; 
            \node[draw=none,fill=none] at (2.1,-0.7){$H$}  ; 
         \node[draw=none,fill=none] at (-1.2,0.7){$q_{i}$}  ; 
        \node[draw=none,fill=none] at (-1.2,-0.7){$\overline{q}_{j}$}  ; 
        
        \fill[red]   (0,0)  circle (0.05cm);
	\draw[korange]         (0,0)  circle (0.05cm);
	\fill[red]   (-0.5,-0.5)  circle (0.05cm);
	\draw[korange]         (-0.5,-0.5)  circle (0.05cm);
        \fill[red]   (-0.5,0.5)  circle (0.05cm);
	\draw[korange]         (-0.5,0.5)  circle (0.05cm);
	 \fill[red]   (1,0)  circle (0.05cm);
	\draw[korange]         (1,0)  circle (0.05cm);
        
\end{tikzpicture}
\hspace{0.5cm}
\begin{tikzpicture}[scale=0.90,
					thick,
			level/.style={level distance=3.15cm, line width=0.4mm},
			level 2/.style={sibling angle=60},
			level 3/.style={sibling angle=60},
			level 4/.style={level distance=1.4cm, sibling angle=60}
	]

	\draw[electron2] (0,0) -- (-1,1) ;
	\draw[electron] (0,0) -- (-1,-1) ;
	\draw[boson] (0,0) -- (1,0) ;
	\draw[boson] (1,0) -- (1.8,0.7) ;
	\draw[boson] (1,0) -- (1.8,-0.7) ;
	\draw[boson] (1.8,0.7) -- (1.8,-0.7) ;
	\draw[boson] (1.8,0.7) -- (2.8,1.2) ;
	\draw[higgs,dashed] (1.8,-0.7) -- (2.8,-1.2) ;
	\node[draw=none,fill=none] at (2.4,0){$V,\gamma$}  ; 
        \node[draw=none,fill=none] at (0.5,0.3){$V$}  ; 
          \node[draw=none,fill=none] at (2.9,0.9){$V$}  ; 
          \node[draw=none,fill=none] at (1.4,0.8){$V$}  ; 
          \node[draw=none,fill=none] at (1.4,-0.9){$V$}  ; 
            \node[draw=none,fill=none] at (2.9,-0.9){$H$}  ; 
         \node[draw=none,fill=none] at (-1.2,0.7){$q_{i}$}  ; 
        \node[draw=none,fill=none] at (-1.2,-0.7){$\overline{q}_{j}$}  ; 
        
        \fill[red]   (0,0)  circle (0.05cm);
	\draw[korange]         (0,0)  circle (0.05cm);
	 \fill[red]   (1,0)  circle (0.05cm);
	\draw[korange]         (1,0)  circle (0.05cm);
	 \fill[red]   (1.8,0.7)  circle (0.05cm);
	\draw[korange]         (1.8,0.7)  circle (0.05cm);
	 \fill[red]   (1.8,-0.7)  circle (0.05cm);
	\draw[korange]         (1.8,-0.7)  circle (0.05cm);
        
\end{tikzpicture}
\hspace{0.5cm}
\begin{tikzpicture}[scale=0.90,
					thick,
			level/.style={level distance=3.15cm, line width=0.4mm},
			level 2/.style={sibling angle=60},
			level 3/.style={sibling angle=60},
			level 4/.style={level distance=1.4cm, sibling angle=60}
	]

	\draw[electron2] (0,-0.5) -- (0,0.5) ;
	\draw[electron2] (0,0.5) -- (-1,1) ;
	\draw[electron] (0,-0.5) -- (-1,-1) ;
	\draw[boson] (0,0.5) -- (1,0) ;
	\draw[boson] (0,-0.5) -- (1,0) ;
	\draw[boson] (1,0) -- (2,0) ;
	\draw[boson] (2,0) -- (3,1) ;
	\draw[higgs,dashed] (2,0) -- (3,-1) ;
	\node[draw=none,fill=none] at (-0.5,0){$q_{i}$}  ; 
        \node[draw=none,fill=none] at (0.7,0.6){$V,\gamma$}  ;
         \node[draw=none,fill=none] at (0.7,-0.55){$V$}  ; 
        \node[draw=none,fill=none] at (1.8,0.3){$V$}  ; 
        \node[draw=none,fill=none] at (3.1,0.6){$V$}  ; 
            \node[draw=none,fill=none] at (3.1,-0.7){$H$}  ; 
         \node[draw=none,fill=none] at (-1.2,0.7){$q_{i}$}  ; 
        \node[draw=none,fill=none] at (-1.2,-0.7){$\overline{q}_{j}$}  ; 
        \fill[red]   (0,0.5)  circle (0.05cm);
	\draw[korange]         (0,0.5)  circle (0.05cm);
	\fill[red]   (0,-0.5)  circle (0.05cm);
	\draw[korange]         (0,-0.5)  circle (0.05cm);
        \fill[red]   (1,0)  circle (0.05cm);
	\draw[korange]         (1,0)  circle (0.05cm);
	 \fill[red]   (2,0)  circle (0.05cm);
	\draw[korange]         (2,0)  circle (0.05cm);
        
\end{tikzpicture}

\begin{flushleft}
	Box diagrams:
	\end{flushleft}

\vspace{0.09cm}
\begin{tikzpicture}[scale=0.90,
					thick,
			level/.style={level distance=3.15cm, line width=0.4mm},
			level 2/.style={sibling angle=60},
			level 3/.style={sibling angle=60},
			level 4/.style={level distance=1.4cm, sibling angle=60}
	]

	\draw[electron2] (-0.5,-0.5) -- (-0.5,0.5) ;
	\draw[electron2] (-0.5,0.5) -- (-1.5,1.5) ;
	\draw[electron] (-0.5,-0.5) -- (-1.5,-1.5) ;
	\draw[boson] (-0.5,0.5) -- (0.5,0.5) ;
	\draw[boson] (-0.5,-0.5) -- (0.5,-0.5) ;
	\draw[boson] (0.5,-0.5) -- (0.5,0.5) ;
	\draw[higgs,dashed] (0.5,-0.5) -- (1.5,-1.5) ;
	\draw[boson] (0.5,0.5) -- (1.5,1.5) ;
	\node[draw=none,fill=none] at (-0.8,0){$q_{i}$}  ; 
        \node[draw=none,fill=none] at (0,0.86){$V$}  ;
        \node[draw=none,fill=none] at (0,-0.86){$V$}  ;
         \node[draw=none,fill=none] at (0.8,0){$V$}  ; 
        \node[draw=none,fill=none] at (1.4,-1){$H$}  ; 
        \node[draw=none,fill=none] at (1.4,0.95){$V$}  ; 
         \node[draw=none,fill=none] at (-1.5,0.95){$q_{i}$}  ; 
        \node[draw=none,fill=none] at (-1.5,-0.9){$\overline{q}_{j}$}  ;
        
        \fill[red]   (-0.5,0.5)  circle (0.05cm);
	\draw[korange]         (-0.5,0.5)  circle (0.05cm);
	\fill[red]   (-0.5,0.5)  circle (0.05cm);
	\draw[korange]         (-0.5,0.5)  circle (0.05cm);
        \fill[red]   (0.5,0.5)  circle (0.05cm);
	\draw[korange]         (0.5,0.5)  circle (0.05cm);
         \fill[red]   (0.5,-0.5)  circle (0.05cm);
	\draw[korange]         (0.5,-0.5)  circle (0.05cm);
\end{tikzpicture}
\hspace{0.5cm}
\begin{tikzpicture}[scale=0.90,
					thick,
			level/.style={level distance=3.15cm, line width=0.4mm},
			level 2/.style={sibling angle=60},
			level 3/.style={sibling angle=60},
			level 4/.style={level distance=1.4cm, sibling angle=60}
	]

	\draw[electron2] (-0.5,-0.5) -- (-0.5,0.5) ;
	\draw[electron2] (-0.5,0.5) -- (-1.5,1.5) ;
	\draw[electron] (-0.5,-0.5) -- (-1.5,-1.5) ;
	\draw[boson] (-0.5,0.5) -- (0.5,0.5) ;
	\draw[boson] (-0.5,-0.5) -- (0.5,-0.5) ;
	\draw[electron2,dashed] (0.5,-0.5) -- (0.5,0.5) ;
	\draw[higgs,dashed] (0.5,-0.5) -- (1.5,-1.5) ;
	\draw[boson] (0.5,0.5) -- (1.5,1.5) ;
	\node[draw=none,fill=none] at (-0.8,0){$q_{i}$}  ; 
        \node[draw=none,fill=none] at (0,0.86){$V,\gamma$}  ;
        \node[draw=none,fill=none] at (0,-0.86){$V$}  ;
         \node[draw=none,fill=none] at (0.8,0){$\varphi$}  ; 
        \node[draw=none,fill=none] at (1.4,-1){$H$}  ; 
        \node[draw=none,fill=none] at (1.4,0.95){$V$}  ; 
         \node[draw=none,fill=none] at (-1.5,0.95){$q_{i}$}  ; 
        \node[draw=none,fill=none] at (-1.5,-0.9){$\overline{q}_{j}$}  ; 
        
        \fill[red]   (-0.5,0.5)  circle (0.05cm);
	\draw[korange]         (-0.5,0.5)  circle (0.05cm);
	\fill[red]   (-0.5,0.5)  circle (0.05cm);
	\draw[korange]         (-0.5,0.5)  circle (0.05cm);
        \fill[red]   (0.5,0.5)  circle (0.05cm);
	\draw[korange]         (0.5,0.5)  circle (0.05cm);
         \fill[red]   (0.5,-0.5)  circle (0.05cm);
	\draw[korange]         (0.5,-0.5)  circle (0.05cm);

\end{tikzpicture}
\hspace{0.5cm}
\begin{tikzpicture}[scale=0.90,
					thick,
			level/.style={level distance=3.15cm, line width=0.4mm},
			level 2/.style={sibling angle=60},
			level 3/.style={sibling angle=60},
			level 4/.style={level distance=1.4cm, sibling angle=60}
	]

	\draw[electron2] (-0.5,-0.5) -- (-0.5,0.5) ;
	\draw[electron2] (-0.5,0.5) -- (-1.5,1.5) ;
	\draw[electron] (-0.5,-0.5) -- (-1.5,-1.5) ;
	\draw[boson] (-0.5,0.5) -- (0.5,0.5) ;
	\draw[boson] (-0.5,-0.5) -- (0.5,-0.5) ;
	\draw[electron,dashed] (0.5,-0.5) -- (0.5,0.5) ;
	\draw[higgs,dashed] (0.5,-0.5) -- (1.5,-1.5) ;
	\draw[boson] (0.5,0.5) -- (1.5,1.5) ;
	\node[draw=none,fill=none] at (-0.8,0){$q_{i}$}  ; 
        \node[draw=none,fill=none] at (0,0.86){$V$}  ;
        \node[draw=none,fill=none] at (0,-0.86){$V$}  ;
         \node[draw=none,fill=none] at (0.8,0){$G$}  ; 
        \node[draw=none,fill=none] at (1.4,-1){$H$}  ; 
        \node[draw=none,fill=none] at (1.4,0.95){$V$}  ; 
         \node[draw=none,fill=none] at (-1.5,0.95){$q_{i}$}  ; 
        \node[draw=none,fill=none] at (-1.5,-0.9){$\overline{q}_{j}$}  ; 
        
        \fill[red]   (-0.5,0.5)  circle (0.05cm);
	\draw[korange]         (-0.5,0.5)  circle (0.05cm);
	\fill[red]   (-0.5,0.5)  circle (0.05cm);
	\draw[korange]         (-0.5,0.5)  circle (0.05cm);
        \fill[red]   (0.5,0.5)  circle (0.05cm);
	\draw[korange]         (0.5,0.5)  circle (0.05cm);
         \fill[red]   (0.5,-0.5)  circle (0.05cm);
	\draw[korange]         (0.5,-0.5)  circle (0.05cm);

\end{tikzpicture}
   	\caption{Feynman diagrams corresponding to
the EW virtual corrections to the LO processes.}
\label{fig:dis}
\end{figure}
\begin{figure}[!htb]
	\centering
\input{ewkVerticesTikz}
   	\caption{Feynman diagrams corresponding to
the EW real corrections to the LO processes.}
\label{fig:feynVertices}
\end{figure}

At NLO the corrections to the  Higgs-strahlung process includes QCD corrections (virtual corrections with gluon exchange in the $q\bar q$ vertex, gluon-induced processes, and the emission of an additional gluon) and complete EW corrections (incoming photon corrections, virtual and real EW corrections including the outgoing photon corrections).
The representative Feynman diagrams for the NLO QCD and EW  corrections are shown in Figs.~\ref{fig:dis}, \ref{fig:feynVertices} and \ref{fig:qcdVertices}.
The EW virtual corrections can be classified into vertex and box corrections.
\begin{figure}[!htb]
\begin{flushleft}
	(a)Virtual diagrams:
	\end{flushleft}	
\begin{tikzpicture}[scale=0.90,
					thick,
			level/.style={level distance=3.15cm, line width=0.4mm},
			level 2/.style={sibling angle=60},
			level 3/.style={sibling angle=60},
			level 4/.style={level distance=1.4cm, sibling angle=60}
	]

        \draw[electron2] (0,0) -- (-1.3,1) ;
	\draw[electron]  (0,0) -- (-1.3,-1) ;
	\draw[gluon] (-0.8,-0.6) -- (-0.8,0.6) ;
	\draw[boson] (0,0) -- (1,0) ;
	\draw[boson] (1,0) -- (2,1) ;
	\draw[higgs,dashed] (1,0) -- (2,-1) ;
	\node[draw=none,fill=none] at (-1.2,0){$g$}  ; 
        \node[draw=none,fill=none] at (0.7,0.4){$V$}  ; 
        \node[draw=none,fill=none] at (2.2,0.7){$V$}  ; 
        \node[draw=none,fill=none] at (2.2,-0.7){$H$}  ; 
         \node[draw=none,fill=none] at (-1.3,0.6){$q_{i}$}  ; 
        \node[draw=none,fill=none] at (-1.3,-0.6){$\overline{q}_{j}$}  ; 
	
	\fill[red]   (0,0)  circle (0.05cm);
	\draw[korange]         (0,0)  circle (0.05cm);
	\fill[red]   (1,0)  circle (0.05cm);
	\draw[korange]         (1,0)  circle (0.05cm);
       \fill[red]   (-0.8,-0.6)  circle (0.05cm);
	\draw[korange]         (-0.8,-0.6)  circle (0.05cm);
	\fill[red]   (-0.8,0.6)  circle (0.05cm);
	\draw[korange]         (-0.8,0.6)  circle (0.05cm);
\end{tikzpicture}
\vspace{0.7cm}	

\begin{flushleft}
	(b)Real diagrams:
	\end{flushleft}		

	\begin{tikzpicture}[scale=0.90,
					thick,
			level/.style={level distance=3.15cm, line width=0.4mm},
			level 2/.style={sibling angle=60},
			level 3/.style={sibling angle=60},
			level 4/.style={level distance=1.4cm, sibling angle=60}
	]

	\draw[electron] (1,0) -- (2,-1) ;
	\draw[gluon] (0,0) -- (-1.3,1) ;
	\draw[electron2] (0,0) -- (-1.3,-1) ;
	\draw[electron] (0,0) -- (1,0) ;
	\draw[boson] (1,0) -- (1.7,1) ;
	\draw[boson] (1.7,1) -- (2.5,2) ;
	\draw[higgs,dashed] (1.7,1) -- (2.5,0) ;
	 \node[draw=none,fill=none] at (1.2,0.9){$V$}  ; 
        \node[draw=none,fill=none] at (2.5,1.5){$V$}  ; 
        \node[draw=none,fill=none] at (2.5,0.4){$H$}  ; 
         \node[draw=none,fill=none] at (1.5,-1.1){$\overline{q}_{j}$}  ; 
          \node[draw=none,fill=none] at (-1.3,-0.7){$q_{i}$}  ;
        \node[draw=none,fill=none] at (-1.3,0.6){$g$}  ; 
        \node[draw=none,fill=none] at (0.5,0.45){$q_{i}$}  ;
	
	\fill[red]   (0,0)  circle (0.05cm);
	\draw[korange]         (0,0)  circle (0.05cm);
	\fill[red]   (1,0)  circle (0.05cm);
	\draw[korange]         (1,0)  circle (0.05cm);
\fill[red]   (1.7,1)  circle (0.05cm);
	\draw[korange]         (1.7,1)  circle (0.05cm);
\end{tikzpicture}		
\hspace{0.8cm}
\begin{tikzpicture}[scale=0.90,
					thick,
			level/.style={level distance=3.15cm, line width=0.4mm},
			level 2/.style={sibling angle=60},
			level 3/.style={sibling angle=60},
			level 4/.style={level distance=1.4cm, sibling angle=60}
	]

	\draw[electron] (0,0) -- (0,0.8) ;
	\draw[gluon] (0,0.8) -- (-1.3,1.8) ;
	\draw[electron2] (0,0) -- (-1.3,-1) ;
	\draw[electron] (0,0.8) -- (1,1.8) ;
	\draw[boson] (0,0) -- (1,0) ;
	\draw[boson] (1,0) -- (2,1) ;
	\draw[higgs,dashed] (1,0) -- (2,-1) ;
	\node[draw=none,fill=none] at (1.1,1.4){$q_{j}$}  ; 
        \node[draw=none,fill=none] at (0.7,0.4){$V$}  ; 
        \node[draw=none,fill=none] at (2.2,0.7){$V$}  ; 
        \node[draw=none,fill=none] at (2.2,-0.7){$H$}  ; 
         \node[draw=none,fill=none] at (-1.3,1.35){$g$}  ; 
          \node[draw=none,fill=none] at (-0.3,0.4){$\overline{q}_{j}$}  ;
        \node[draw=none,fill=none] at (-1.3,-0.7){$q_{i}$}  ; 
	
	\fill[red]   (0,0)  circle (0.05cm);
	\draw[korange]         (0,0)  circle (0.05cm);
	\fill[red]   (1,0)  circle (0.05cm);
	\draw[korange]         (1,0)  circle (0.05cm);
\fill[red]   (0,0.8)  circle (0.05cm);
	\draw[korange]         (0,0.8)  circle (0.05cm);
\end{tikzpicture}			
\hspace{0.8cm}		
\begin{tikzpicture}[scale=0.90,
					thick,
			level/.style={level distance=3.15cm, line width=0.4mm},
			level 2/.style={sibling angle=60},
			level 3/.style={sibling angle=60},
			level 4/.style={level distance=1.4cm, sibling angle=60}
	]

	\draw[electron2] (0,0) -- (-1.3,1) ;
	\draw[gluon] (0,-1) -- (1,-1.7) ;
	\draw[electron] (0,0) -- (0,-1) ;
	\draw[electron] (0,-1) -- (-1.3,-2) ;
	\draw[boson] (0,0) -- (1,0) ;
	\draw[boson] (1,0) -- (2,1) ;
	\draw[higgs,dashed] (1,0) -- (2,-1) ;
	\node[draw=none,fill=none] at (-0.3,-0.5){$\overline{q}_{j}$}  ; 
        \node[draw=none,fill=none] at (0.7,0.4){$V$}  ; 
        \node[draw=none,fill=none] at (2.2,0.7){$V$}  ; 
        \node[draw=none,fill=none] at (2.2,-0.7){$H$}  ; 
         \node[draw=none,fill=none] at (1,-1.2){$g$}  ; 
          \node[draw=none,fill=none] at (-1.3,0.6){$q_{i}$}  ;
        \node[draw=none,fill=none] at (-1.3,-1.6){$\overline{q}_{j}$}  ; 
	
	\fill[red]   (0,0)  circle (0.05cm);
	\draw[korange]         (0,0)  circle (0.05cm);
	\fill[red]   (1,0)  circle (0.05cm);
	\draw[korange]         (1,0)  circle (0.05cm);
\fill[red]   (0,-1)  circle (0.05cm);
	\draw[korange]         (0,-1)  circle (0.05cm);
\end{tikzpicture}
	\caption{Feynman diagrams corresponding to
the NLO QCD corrections to the LO processes.  }
	\label{fig:qcdVertices}
\end{figure}

In this section, we present numerical results for total and differential cross sections  for associated $W^{\pm}H$ ($W^+\to l^+\nu_l$ and $W^-\to l^-\bar\nu_l$, where $l = e, \mu$) and ZH ($Z\to l^+l^-$) production with  NNPDF2.3qed, NNPDF3.0qed, MRS2004qed, CT14QEDinc, and  LUXqed photon  PDFs at the LHC at 13 TeV including the NLO QCD and EW corrections.

\subsection{Input Parameters and Cuts}\label{cuts}

In our numerical analysis of the VH processes ($V= W^\pm, Z$) we use the same VH input parameters and cuts
as in Ref. \cite{2016-008cc},

\begin{eqnarray}
G_{F} = 1.166378\times10^{-5} GeV^{-2}, \qquad\qquad\qquad\nonumber
\alpha_{s}(M_{Z}) = 0.11801,            \qquad \nonumber  \\
\Gamma_{W}=2.08430~GeV,                  \qquad\qquad\qquad \nonumber
\Gamma_{Z}=2.49427~GeV,                  \qquad \nonumber\\
 M_{H}=125~GeV,                          \qquad\qquad \nonumber
 M_{W} = 80.35797~GeV,                   \qquad\qquad\nonumber
 M_{Z} = 91.15348~GeV ,                  \nonumber\\
 m_{e} = 0.510998910 MeV,               \qquad\qquad \nonumber
 m_{\mu}= 0.105658 GeV,                  \qquad\qquad \nonumber
 m_{t}= 172.5 GeV
\end{eqnarray}

The electromagnetic coupling is fixed in the $G_{F}$-scheme,

\begin{equation}
 \alpha_{G_F}=\dfrac{\sqrt{2}G_{F}M_{W}^{2}}{\pi}\sin^{2}\theta_{W}
\end{equation}

and the
weak mixing angle is defined in the on-shell scheme,
\begin{equation}
 \sin^{2}\theta_{W}=1-M_W^2/M_Z^2
\end{equation}

Both the factorization and the renormalization scales are set to the value of the  invariant mass of the VH system,
\begin{equation}
\mu=\mu_R=\mu_{F} = M_{VH}, \;\; \; M^2_{VH} \equiv (p_{V} + p_{H})^2.
\end{equation}

We also provide results for the total cross section and  differential cross section
 in the same kinematic region as in Ref.~\cite{2016-008cc},

 \begin{equation}
\label{taggingjets}
p_{T_{l}} > 15 GeV,  \qquad  |y_{l}|<2.5,
\end{equation}
where $p_{T_{l}}$ is the transverse momentum of the lepton and $y_{l}$ is
its rapidity.
For $Z(\to l^+ l^-)H$ production the invariant mass of the two
leptons is in the range
\begin{equation}
\label{taggingjets}
75 GeV < M_{ll} < 105 GeV.
\end{equation}

\subsection{ Total Cross Sections}

First we study the effect of the NLO corrections on the LO cross sections  without any cuts on the final-state charged leptons using the SM input parameters given in Section~\ref{cuts}.
The total NLO VH cross sections $\sigma^{VH}_{NLO}$ and relative corrections $\delta_{QCD/EW}$ are calculated according to
\begin{equation}
\label{taggingjets}
\sigma^{VH}_{NLO}=\sigma^{VH}_{LO}(1+\delta_{EW}+\delta_{QCD})+\sigma_{\gamma},
\end{equation}
and
\begin{equation}
\label{taggingjets2}
\delta_{QCD/EW}=\dfrac{\sigma_{QCD/EW}}{\sigma^{VH}_{LO}} \times100\% .
\end{equation}
For the LO calculations, strictly speaking, one should use LO PDFs, but for convenience, we  used NLO PDFs.
In Tables \ref{ta:xsection_nocuts}, \ref{ta:xsection_nocuts2}, and \ref{ta:xsection_nocuts3}
we list the LO and NLO total cross sections ($\sigma_{\mathrm{LO}}$, $\sigma_{\mathrm{NLO}}$)
including NLO QCD and EW corrections ($\delta_{\mathrm{QCD}}$, $\delta_{\mathrm{EW}}$),
 and photon-induced cross sections ($\sigma_{\gamma}$, $\sigma_{\gamma\; PR}$)  with PDF uncertainties for the Higgs-strahlung processes  $pp \to W^\pm /Z + H  \to l\nu_l \//l^+l^- +H$ with NNPDF2.3qed, NNPDF3.0qed, MRS2004qed, CT14QEDinc, and  LUXqed photon  PDFs and  Higgs-boson mass $M_{H}$=125 GeV at the LHC at $\sqrt{s}$=13 TeV.  The photon-induced cross section is only significant for WH production. Note that we did not apply reweighting to the LUXqed photon PDFs since the PDF uncertainty of the  photon-induced cross section is already small.
 We see from Tables \ref{ta:xsection_nocuts} and \ref{ta:xsection_nocuts2} that the  photon-induced cross section $\sigma_{\gamma}$ and uncertainties depend on the choice of the photon PDFs. Especially,
 the photon PDF uncertainties decrease significantly as we change from  NNPDF2.3qed and NNPDF3.0qed to MRST2004qed, CT14QEDinc, and to LUXqed.
 The NLO QCD corrections  for various photon PDFs are of the order of $+17.0\%$  for total cross sections.
The EW corrections are insensitive to the choice of the photon  PDFs, and are about $-7.0\%$.
The reweighting PDF uncertainties  of the photon-induced cross section $\sigma_{\gamma\; PR}$
 are reduced significantly after including the CMS data to reweight the NNPDF2.3qed, NNPDF3.0qed, MRS2004qed, and CT14QEDinc photon PDFs, while the corresponding central predicted cross sections are unchanged.
 From  Table \ref{ta:xsection_nocuts} ($p p \to H+ W^+(\to l^+\nu_l)$),
the relative errors $\Delta\sigma_\gamma/\sigma_\gamma$ for the photon-induced cross sections reduce from  88.7\%  and 89.5\% in the NNPDF2.3qed and NNPDF3.0qed predictions to 5.5\% and 8.2\% in  the reweighting NNPDF2.3qed and reweighting NNPDF3.0qed predictions;
the relative errors $\Delta\sigma_\gamma/\sigma_\gamma$  reduce from  23.9\%  in the MRST2004qed predictions to 7.6\% in the reweighting MRST2004qed predictions;
the relative errors $\Delta\sigma_\gamma/\sigma_\gamma$  reduce from  13.4\% in  the CT14QEDinc predictions to 7.1\%  in the reweighting CT14QEDinc predictions.
\begin{table}[!htb]
\def\phm{\phantom{-}}
\def\phn{\phantom{0}}
\centerline{
\begin{tabular}{|c|c|c|c|c|c|}
\hline
PDF & NNPDF2.3qed & NNPDF3.0qed & MRST2004qed & CT14QEDinc & LUXqed   \\
\hline
$\sigma_{LO}\ [fb]$
& $79.06\pm1.15$  
& $78.45\pm1.17$  
& $78.75\pm0.03$  
& $78.75\pm0.07 $ 
& $80.59\pm0.13$  
\\
$\sigma_{NLO}\ [fb]$
& $93.66\pm5.70$  
& $92.30\pm5.56$  
& $92.43\pm0.61 $ 
& $90.48\pm0.27 $ 
& $93.36\pm0.15 $ 
\\
$\sigma_{\gamma}\ [fb]$
& $6.01\pm5.33$  
& $5.94\pm5.32$  
& $4.46\pm0.65$  
& $4.47\pm0.36$  
& $4.49\pm0.006$  
\\
$\delta_{QCD}\ [\%]$
& $17.53\pm0.04$  
& $17.47\pm0.05$  
& $17.75\pm0.005$  
& $17.39\pm0.001$  
& $17.68\pm0.008$  
\\
$\delta_{EW}\ [\%]$
& $-7.41\pm0.008$  
& $-7.43\pm0.009$  
& $-7.42\pm0.006$  
& $-7.41\pm0.006$  
& $-7.41\pm0.001$  
\\
\hline
$\sigma_{\gamma\; PR}\ [fb]$
& $6.01\pm0.33$  
& $5.94\pm0.49$  
& $4.46\pm0.21$  
& $4.47\pm0.19$  
& $4.49\pm0.006$  
\\
\hline
\end{tabular}
}
\caption{
The first and second rows  are  the total cross sections in LO and NLO;
the third row is the photon-induced process cross sections; and
the fourth and fifth rows are the NLO QCD and EW corrections and corresponding PDF uncertainties (68\% CL) for the process $p p \to H+ W^+(\to l^+\nu_l)$ with various PDFs at the LHC at 13 TeV and Higgs boson mass $ M_H = 125$ GeV without kinematic cuts.
The last row is the photon-induced process cross sections and PDF uncertainties for the various re-weighted PDFs.
}\label{ta:xsection_nocuts}
\end{table}
From Table \ref{ta:xsection_nocuts2} ($pp \to H+ W^-(\to l^-\bar \nu_l)$) we see that
the relative errors $\Delta\sigma_\gamma/\sigma_\gamma$ of the photon-induced cross sections reduce from  96.6\%  and 95.4\% in the NNPDF2.3qed and NNPDF3.0qed predictions to  5.7\% and 9.5\% in  the reweighting NNPDF2.3qed and reweighting NNPDF3.0qed predictions;
the relative errors $\Delta\sigma_\gamma/\sigma_\gamma$  reduce from  24.5\%  in the MRST2004qed predictions to 7.6\% in the reweighting MRST2004qed predictions; and
the relative errors $\Delta\sigma_\gamma/\sigma_\gamma$  reduce from  11.8\% in  the CT14QEDinc predictions to 6.3\%  in the reweighting CT14QEDinc predictions.
\begin{table}[!htb]
\def\phm{\phantom{-}}
\def\phn{\phantom{0}}
\centerline{
\begin{tabular}{|c|c|c|c|c|c|}
\hline
PDF & NNPDF2.3qed & NNPDF3.0qed & MRST2004qed & CT14QEDinc & LUXqed   \\
\hline
$\sigma_{LO}\ [fb]$
& $50.41\pm0.73 $ 
& $49.77\pm0.85 $ 
& $49.35\pm0.02 $ 
& $49.73\pm0.06 $ 
& $51.36\pm0.09 $ 
\\
$\sigma_{NLO}\ [fb]$
& $59.41\pm4.27$  
& $58.65\pm4.23$  
& $57.78\pm0.41$  
& $56.99\pm0.14$  
& $59.35\pm0.11$  
\\
$\sigma_{\gamma}\ [fb]$
& $4.19\pm4.05$  
& $4.18\pm3.99$  
& $2.89\pm0.43$  
& $2.86\pm0.20$  
& $2.98\pm0.004$  
\\
$\delta_{QCD}\ [\%]$
& $16.79\pm0.06$  
& $16.73\pm0.05$  
& $17.05\pm0.005$  
& $16.79\pm0.002$  
& $17.02\pm0.008$  
\\
$\delta_{EW}\ [\%]$
& $-7.26\pm0.007$  
& $-7.30\pm0.0006$  
& $-7.28\pm0.0005$  
& $-7.28\pm0.006$  
& $-7.26\pm0.001$  
\\
\hline
$\sigma_{\gamma\; PR}\ [fb]$
& $4.19\pm0.24$  
& $4.18\pm0.40$  
& $2.89\pm0.13$  
& $2.86\pm0.11$  
& $2.98\pm0.004$  
\\
\hline
\end{tabular}
 }
\caption{
The first and second rows  are  the total cross sections in LO and NLO;
the third row is the photon-induced process cross sections; and
the fourth and fifth rows are the NLO QCD and EW corrections and corresponding PDF uncertainties (68\% CL)
for the process $pp \to H+ W^-(\to l^-\bar\nu_l)$ with various PDFs at the LHC at 13~TeV and Higgs boson mass $ M_H = 125$ GeV without kinematic cuts.
 The last row is the photon-induced process cross sections and PDF uncertainties for the various re-weighted PDFs.
 }
\label{ta:xsection_nocuts2}
\end{table}
From Table \ref{ta:xsection_nocuts3} ($pp \to H+ Z(\to l^+l^-)$) we see that
the  photon-induced cross section contributions is small for various photon PDFs;
the NLO QCD corrections are of the order of $+17\%$  for total cross sections;
the EW corrections are insensitive to the choice of the PDF set, and are about $-5\%$;
the PDF uncertainties  of the photon-induced cross sections $\sigma_{\gamma\; PR}$
 are reduced significantly after including the CMS data to reweight the various photon PDFs, while the average predicted cross sections are unchanged.
For example, the relative errors $\Delta\sigma_\gamma/\sigma_\gamma$ of the photon-induced cross sections reduce from   76.2\%  and 86.5\% in the NNPDF2.3qed and NNPDF3.0qed predictions to   4.7\% and 8.1\%; in  the reweighting NNPDF2.3qed and reweighting NNPDF3.0qed predictions;
the relative errors $\Delta\sigma_\gamma/\sigma_\gamma$  reduce from  17.2\%   in the MRST2004qed predictions to 5.1\% in the reweighting MRST2004qed predictions; and
the relative errors $\Delta\sigma_\gamma/\sigma_\gamma$  reduce from 10.1\% in  the CT14QEDinc predictions to 3.7\%  in the reweighting CT14QEDinc predictions.
\begin{table}[!htb]
\def\phm{\phantom{-}}
\def\phn{\phantom{0}}
\centerline{
\begin{tabular}{|c|c|c|c|c|c|}
\hline
PDF & NNPDF2.3qed & NNPDF3.0qed & MRST2004qed & CT14QEDinc & LUXqed   \\
\hline
$\sigma_{LO}\ [fb]$
& $21.77\pm0.31$  
& $21.58\pm0.32$  
& $21.72\pm0.0006 $ 
& $21.68\pm0.02$  
& $22.26\pm0.03$  
\\
$\sigma_{NLO}\ [fb]$
& $24.74\pm0.56 $ 
& $24.46\pm0.51 $ 
& $24.65\pm0.02 $ 
& $24.47\pm0.006 $ 
& $25.19\pm0.04 $ 
\\
$\sigma_{\gamma}\ [fb]$
& $0.42\pm0.32$  
& $0.37\pm0.32$  
& $0.29\pm0.03$  
& $0.30\pm0.02$  
& $0.29\pm0.0003$  
\\
$\delta_{QCD}\ [\%]$
& $16.82\pm0.05$  
& $16.71\pm0.05$  
& $17.02\pm0.005$  
& $16.79\pm0.002$  
& $17.00\pm0.009$  
\\
$\delta_{EW}\ [\%]$
& $-5.14\pm0.005$  
& $-5.13\pm0.005$  
& $-5.10\pm0.002$  
& $-5.16\pm0.0002$  
& $-5.15\pm0.0008$  
\\
\hline
$\sigma_{\gamma\; PR}\ [fb]$
& $0.42\pm0.02$  
& $0.37\pm0.03$  
& $0.29\pm0.006$  
& $0.30\pm0.006$  
& $0.29\pm0.0003$  
\\
\hline
\end{tabular}
}
\caption{
The first and second rows  are  the total cross sections in LO and NLO; the third row is the photon-induced process cross sections; and the fourth and fifth rows are the NLO QCD and EW corrections and corresponding PDF uncertainties (68\% CL) for the process $\,p\ p \to H+ Z(\to l^+l^-)$ with various PDFs at the LHC at 13~TeV and Higgs boson mass $ M_H = 125$ GeV. The last row is the photon-induced process cross sections and corresponding PDF uncertainties for the various re-weighted PDFs.
 }
\label{ta:xsection_nocuts3}
\end{table}

\subsection{Fiducial  Cross Section}
Apart from the total cross section without any  cuts, in this subsection
we consider the total cross section after applying cuts to the charged leptons.
In Tables \ref{ta:xsection_withcuts4}, \ref{ta:xsection_withcuts5}, and \ref{ta:xsection_withcuts6}
we list the LO and NLO total cross sections ($\sigma_{\mathrm{LO}}$, $\sigma_{\mathrm{NLO}}$)
including NLO QCD and EW corrections ($\delta_{\mathrm{QCD}}$, $\delta_{\mathrm{EW}}$),
and photon-induced cross sections ($\sigma_{\gamma}$, $\sigma_{\gamma\; PR}$) with PDF uncertainties for the Higgs-strahlung processes  $pp \to W^\pm /Z + H  \to l\nu_l \//l^+l^- +H$   with kinematical cuts  in Section \ref{cuts}  with NNPDF2.3qed, NNPDF3.0qed, MRS2004qed, CT14QEDinc, and  LUXqed photon  PDFs and  Higgs boson mass $M_{H}$=125 GeV at the LHC with $\sqrt{s}$=13 TeV.
The LO and NLO total cross sections are reduced by the kinematical cuts.
The NLO QCD corrections for VH  do not depend on the cuts on the final state charged lepton, but the  EW corrections and the  photon-induced cross section do depend on them.
The photon-induced cross section is only significant for WH production.  It is about  $10^{-3}$ fb for the $pp \to H+ Z(\to l^+l^-)$ process, as we can see from Table \ref{ta:xsection_withcuts6}.
Note that we did not apply reweighting to the LUXqed photon PDFs, since the PDF uncertainty of the  photon-induced cross section is already small.
The NLO QCD corrections  for various photon PDFs are of the order of $+17.0\%$  for total cross sections.
The EW corrections are insensitive to the choice of the photon  PDFs, and are about $-8.0\%$ for $W^+H$, $-7.0\%$ $W^-H$ and  $-9.0\%$ for  $ZH$ with leptonic decays.
We see from Tables \ref{ta:xsection_withcuts4} and \ref{ta:xsection_withcuts5}  that the  photon-induced cross sections $\sigma_{\gamma}$ and uncertainties depend on the choice of the photon PDFs. Especially,
 the photon PDF uncertainties decrease significantly as we change from  NNPDF2.3qed and NNPDF3.0qed to MRST2004qed, CT14QEDinc, and to LUXqed.
The reweighting PDF uncertainties  of the photon-induced cross sections $\sigma_{\gamma\; PR}$
 are reduced significantly after including the CMS data to reweight the NNPDF2.3qed, NNPDF3.0qed, MRS2004qed, and CT14QEDinc photon PDFs, while the corresponding central predicted cross sections are unchanged.
From Table \ref{ta:xsection_withcuts4} ($pp \to H+ W^+(\to l^+\nu_l)$) we see that
the relative errors $\Delta\sigma_\gamma/\sigma_\gamma$ for the photon-induced cross sections reduce from  55.5\% and 58.7\%  in the NNPDF2.3qed and NNPDF3.0qed predictions to 10.5\% and 7.2\% in  the reweighting NNPDF2.3qed and reweighting NNPDF3.0qed predictions;
the relative errors $\Delta\sigma_\gamma/\sigma_\gamma$  reduce from 21.6\% in the MRST2004qed predictions to 6.6\% in the reweighting MRST2004qed predictions;
the relative errors $\Delta\sigma_\gamma/\sigma_\gamma$  reduce from  10.6\% in  the CT14QEDinc predictions to 5.5\% in the reweighting CT14QEDinc predictions.
\begin{table}[!htb]
\def\phm{\phantom{-}}
\def\phn{\phantom{0}}
\centerline{
\begin{tabular}{|c|c|c|c|c|c|}
\hline
PDF & NNPDF2.3qed & NNPDF3.0qed & MRST2004qed & CT14QEDinc & LUXqed   \\
\hline
$\sigma_{LO}\ [fb]$
& $65.09\pm0.96 $ 
& $64.61\pm0.98 $ 
& $64.59\pm0.03 $ 
& $64.78\pm0.07 $ 
& $66.33\pm0.11 $ 
\\
$\sigma_{NLO}\ [fb]$
& $73.64\pm1.80 $ 
& $72.94\pm1.90 $ 
& $73.58\pm0.29 $ 
& $72.95\pm0.09 $ 
& $75.05\pm0.12 $ 
\\
$\sigma_{\gamma}\ [fb]$
& $2.56\pm1.42$  
& $2.35\pm1.38$  
& $2.40\pm0.32$  
& $2.54\pm0.16$  
& $2.51\pm0.003$  
\\
$\delta_{QCD}\ [\%]$
& $17.41\pm0.04$  
& $17.41\pm0.04$  
& $17.69\pm0.003$  
& $17.36\pm0.02$  
& $17.57\pm0.007$  
\\
$\delta_{EW}\ [\%]$
& $-8.27\pm0.01$  
& $-8.18\pm0.008$  
& $-8.24\pm0.0002$  
& $-8.26\pm0.01$  
& $-8.21\pm0.001$  
\\
\hline
$\sigma_{\gamma\; PR}\ [fb]$
& $2.56\pm0.27$  
& $2.35\pm0.17$  
& $2.40\pm0.10$  
& $2.54\pm0.09$  
& $2.51\pm0.003$  
\\
\hline
\end{tabular}
}
\caption{
The first and second rows  are  the fiducial cross sections in LO and NLO;
the third row is the photon-induced process fiducial cross sections; and
the fourth and fifth rows are the NLO QCD and EW corrections and corresponding PDF uncertainties (68\% CL)
for the process $pp \to H+ W^+(\to l^+\nu_l)$ with various PDFs at the LHC at 13~TeV and Higgs boson mass $ M_H = 125$ GeV without kinematic cuts.
 The last row is the photon-induced process fiducial cross sections and PDF uncertainties with various re-weighted PDFs.
 }
\label{ta:xsection_withcuts4}
\end{table}

From Table~\ref{ta:xsection_withcuts5} ($pp \to H+ W^-(\to l^-\bar\nu_l)$) we see that
the relative errors $\Delta\sigma_\gamma/\sigma_\gamma$ for the photon-induced cross section reduce from  57.6\%  and 62.4\%  in the NNPDF2.3qed and NNPDF3.0qed predictions to 10.3\% and 7.4\% in  the reweighting NNPDF2.3qed and reweighting NNPDF3.0qed predictions;
the relative errors $\Delta\sigma_\gamma/\sigma_\gamma$  reduce from 22.2\% in the MRST2004qed predictions to 6.9\% in the reweighting MRST2004qed predictions; and
the relative errors $\Delta\sigma_\gamma/\sigma_\gamma$  reduce from  11.3\% in  the CT14QEDinc predictions to 5.9\% in the reweighting CT14QEDinc predictions.
\begin{table}[!htb]
\def\phm{\phantom{-}}
\def\phn{\phantom{0}}
\centerline{
\begin{tabular}{|c|c|c|c|c|c|}
\hline
PDF & NNPDF2.3qed & NNPDF3.0qed & MRST2004qed & CT14QEDinc & LUXqed   \\
\hline
$\sigma_{LO}\ [fb]$
& $37.70\pm0.52 $ 
& $37.25\pm0.62 $ 
& $37.23\pm0.18  $
& $ 37.33\pm0.04 $ 
& $38.46\pm0.07 $ 
\\
$\sigma_{NLO}\ [fb]$
& $42.96\pm1.16 $ 
& $42.29\pm1.25 $ 
&$ 42.62\pm0.18 $ 
&$ 42.22\pm0.04 $ 
& $43.74\pm0.08 $ 
\\
$\sigma_{\gamma}\ [fb]$
& $1.65\pm0.95$  
& $1.49\pm0.93$  
& $1.44\pm0.19$  
& $1.51\pm0.11$  
& $1.54\pm0.002$  
\\
$\delta_{QCD}\ [\%]$
& $17.42\pm0.06$  
& $17.38\pm0.05$  
& $17.66\pm0.005$  
& $17.42\pm0.004$  
& $17.60\pm0.007$  
\\
$\delta_{EW}\ [\%]$
& $-7.96\pm0.008$  
& $-7.87\pm0.008$  
& $-7.95\pm0.0002$  
& $-7.94\pm0.006$  
& $-7.89\pm0.001$  
\\
\hline
$\sigma_{\gamma\; PR}\ [fb]$
& $1.65\pm0.17$  
& $1.49\pm0.11$  
& $1.44\pm0.06$  
& $1.51\pm0.05$  
& $1.54\pm0.002$  
\\
\hline
\end{tabular}
}
\caption{
The first and second rows  are  the fiducial cross sections in LO and NLO;
the third row is the photon-induced process fiducial cross sections;
the fourth and fifth rows are the NLO QCD and EW corrections and corresponding PDF uncertainties (68\% CL)
for the process $pp \to H+ W^-(\to l^-\bar\nu_l)$ with various PDFs at the LHC at 13~TeV and Higgs boson mass $ M_H = 125$ GeV without kinematic cuts.
 }
\label{ta:xsection_withcuts5}
\end{table}

\begin{table}[!htb]
\def\phm{\phantom{-}}
\def\phn{\phantom{0}}
\centerline{
\begin{tabular}{|c|c|c|c|c|c|}
\hline
PDF & NNPDF2.3qed & NNPDF3.0qed & MRST2004qed & CT14QEDinc & LUXqed   \\
\hline
$\sigma_{LO}\ [fb]$
& $13.50\pm0.18  $
& $13.64\pm0.20  $
& $13.48\pm0.006  $
& $13.44\pm0.01  $
& $14.05\pm0.02  $
\\
$\sigma_{NLO}\ [fb]$
& $14.63\pm0.20 $ 
& $14.76\pm0.22 $ 
& $14.65\pm0.006 $ 
& $14.58\pm0.02 $ 
& $15.24\pm0.02 $ 
\\
$\sigma_{\gamma}\ [fb]$
& $1\times10^{-3}$  
& $6\times10^{-3}$  
& $2\times10^{-3}$  
& $2\times10^{-3}$  
& $7\times10^{-3}$  
\\
$\delta_{QCD}\ [\%]$
& $17.37\pm0.05$  
& $17.30\pm0.04$  
& $17.67\pm0.003$  
& $17.45\pm0.003$  
& $17.53\pm0.009$  
\\
$\delta_{EW}\ [\%]$
& $-9.01\pm0.008$  
& $-9.15\pm0.008$  
& $-9.01\pm0.0002$  
& $-8.99\pm0.004$  
& $-9.15\pm0.001$  
\\
\hline
\end{tabular}
}
\caption{
The first and second rows  are  the fiducial cross sections in LO and NLO; the third row is the photon-induced process fiducial cross sections;  and
the fourth and fifth rows are the NLO QCD and EW corrections and corresponding PDF uncertainties (68\% CL)
for the process $\,p\ p \to H+ Z(\to l^+l^-)$ with various PDFs at the LHC at 13~TeV and Higgs boson mass $ M_H = 125$ GeV.
 }
\label{ta:xsection_withcuts6}
\end{table}

\subsection{Fiducial Differential Cross Section}

\begin{figure}[!htbp]
\subfigure{
\begin{minipage}{7cm}
\includegraphics[width=8cm,height=7cm]{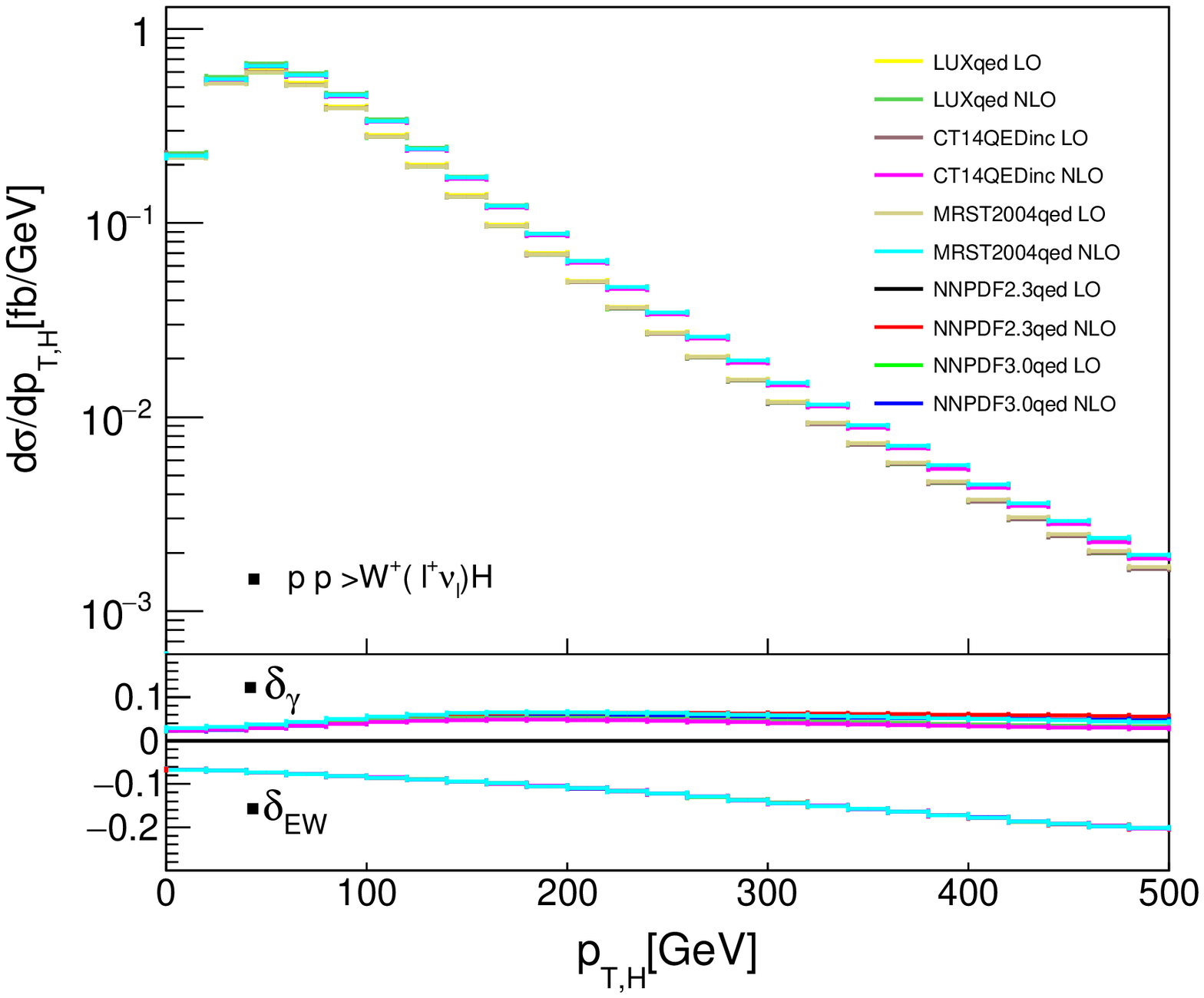}
\end{minipage}
}
\subfigure{
\begin{minipage}{7cm}
\includegraphics[width=8cm,height=7cm]{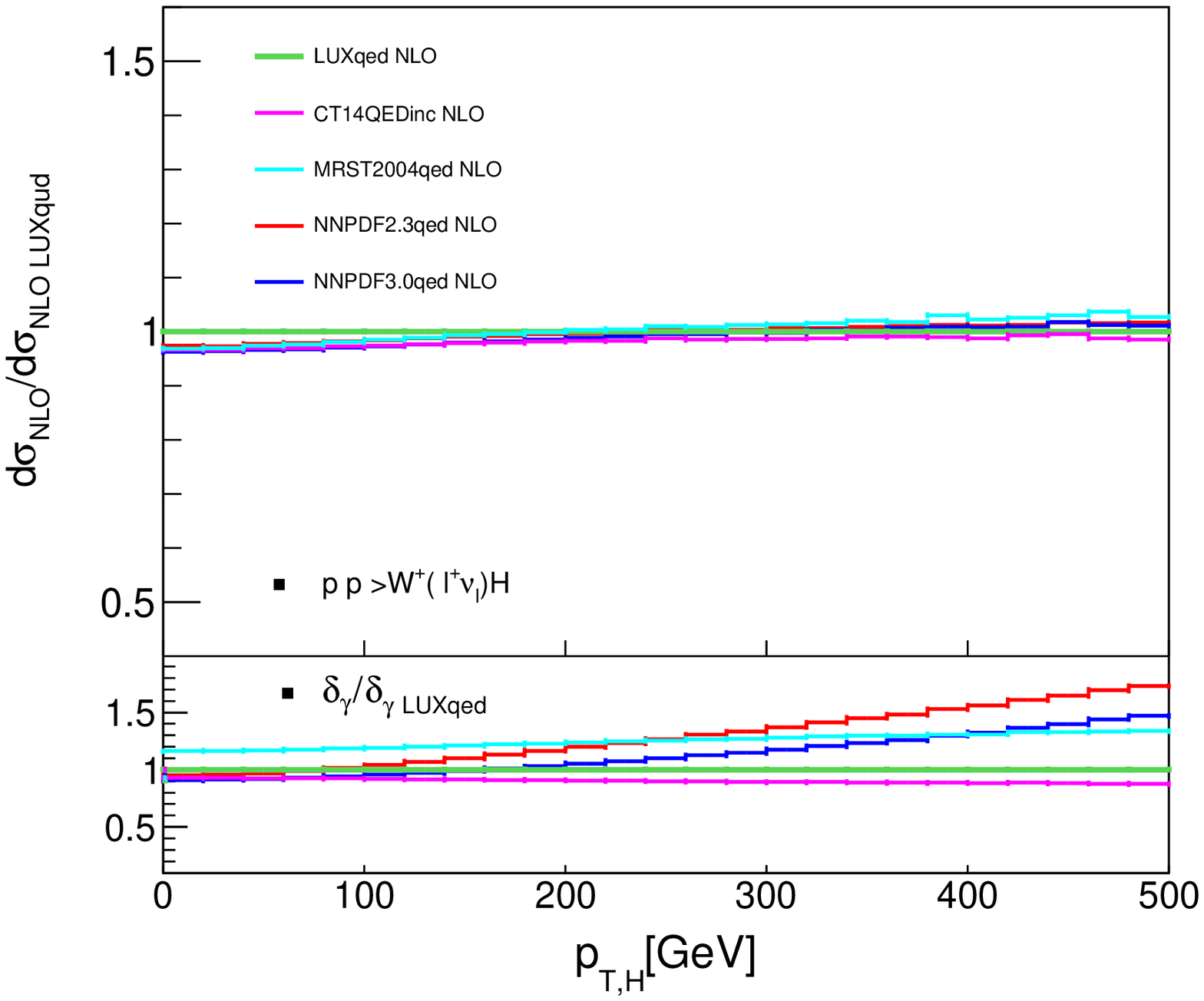}
\end{minipage}
}
\caption{
Fiducial differential cross section distributions (top left panel) both at LO (lower curves) and NLO (upper curves) and ratio of $ d\sigma_{NLO}/d\sigma_{NLO LUXqed}$ (top right panel)  of the Higgs boson in  $W^+( l^+\nu_l)H$ production after applying the selection cuts in Section \ref{cuts} at the LHC at 13 TeV for the various PDFs. The bottom left panel shows the electroweak $\delta_{EW}$ and photon corrections $\delta_\gamma =d\sigma_\gamma/d\sigma_{LO}$. The bottom right panel is the ratio of $\delta_\gamma/\delta_{\gamma LUXqed}$.
}
\label{fig-2-22}
\end{figure}

\begin{figure}[!htbp]
\subfigure{
\begin{minipage}{7cm}
\includegraphics[width=8cm,height=7cm]{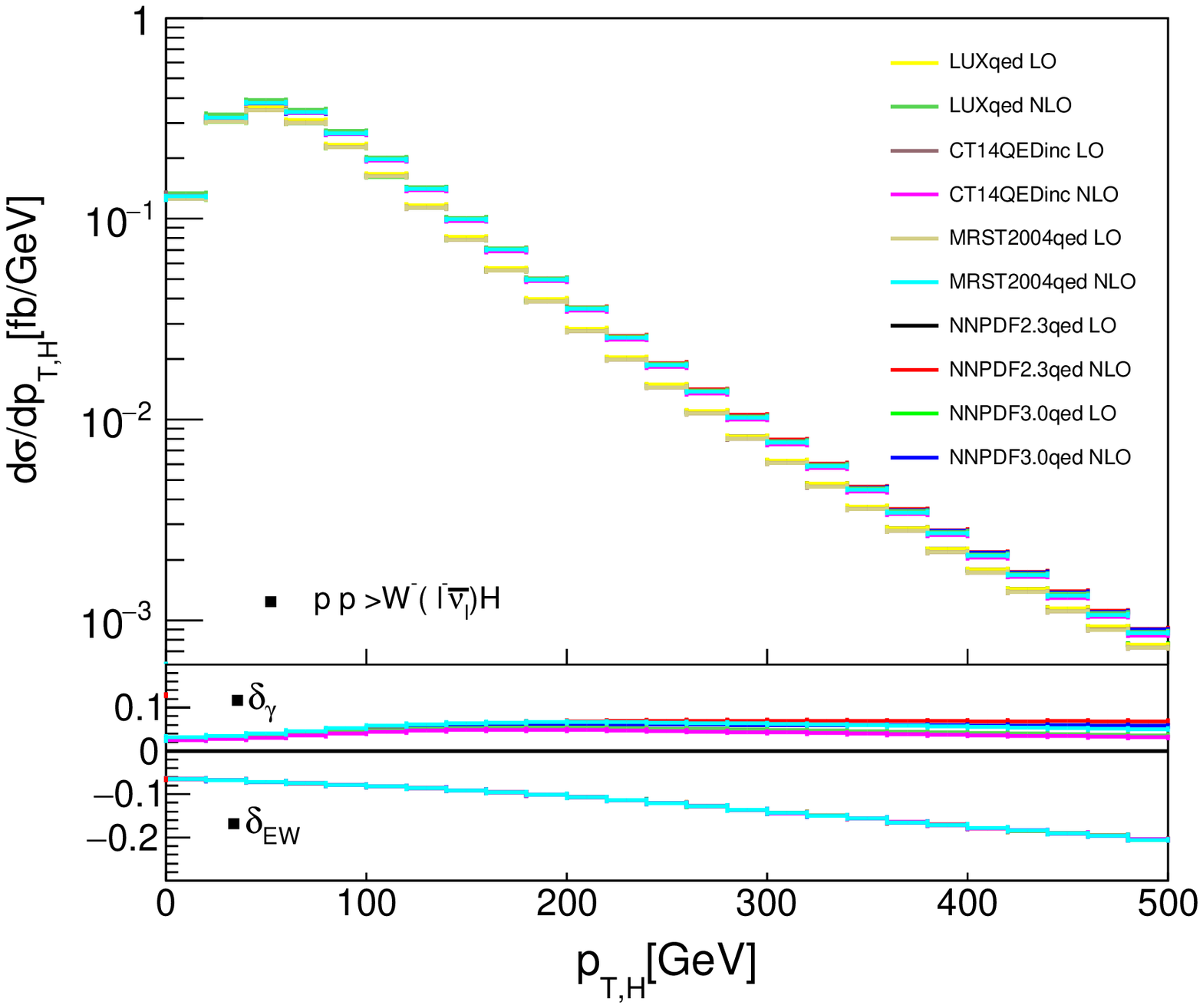}
\end{minipage}
}
\subfigure{
\begin{minipage}{7cm}
\includegraphics[width=8cm,height=7cm]{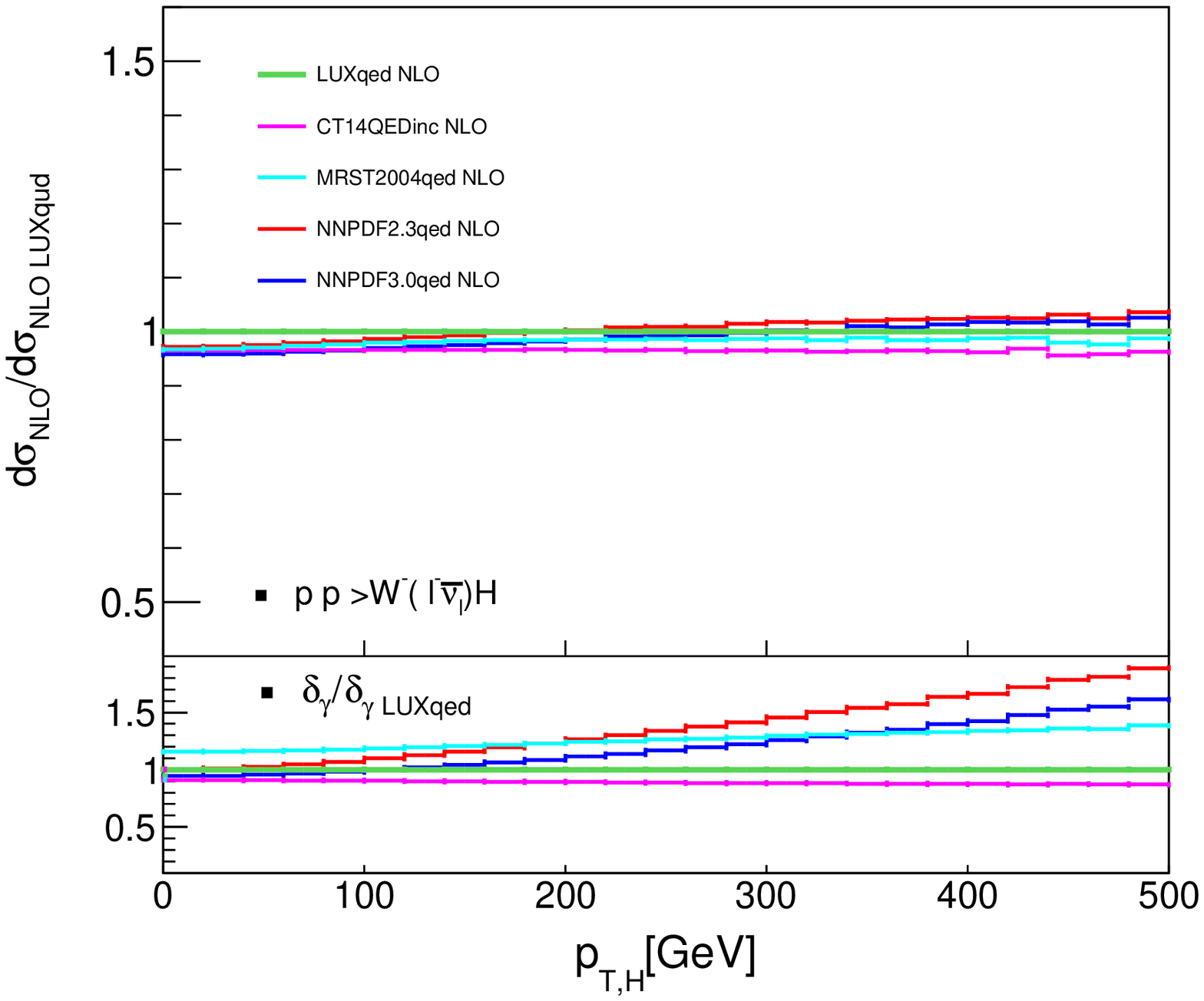}
\end{minipage}
}
\caption{Differential cross section distributions (top left panel) both at LO (lower curves) and NLO (upper curves) and ratio of $ d\sigma_{NLO}/d\sigma_{NLO LUXqed}$ (top right  panel)  of the Higgs boson in $W^-( l^-\bar\nu_l)H$ production after applying the selection cuts in section \ref{cuts} at the LHC at 13 TeV for the various PDFs.
 The bottom left  panel shows the electroweak $\delta_{EW}$ and photon corrections $\delta_\gamma =d\sigma_\gamma/d\sigma_{LO}$. The bottom right  panel is the ratio of $\delta_\gamma/\delta_{\gamma LUXqed}$.
 }
\label{fig-3-33}
\end{figure}

\begin{figure}[!htbp]
\subfigure{
\begin{minipage}{7cm}
\includegraphics[width=8cm,height=7cm]{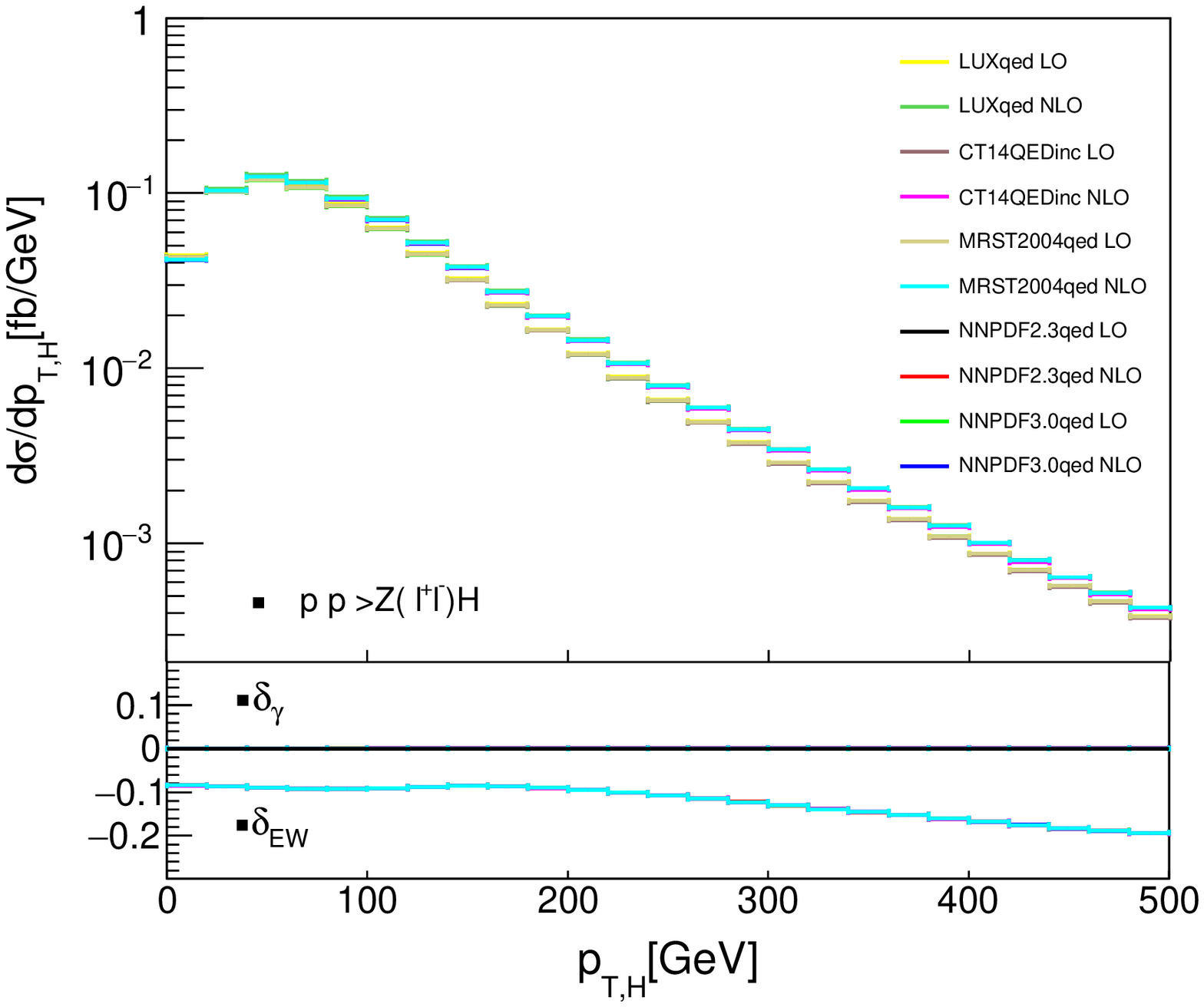}
\end{minipage}
}
\subfigure{
\begin{minipage}{7cm}
\includegraphics[width=8cm,height=7cm]{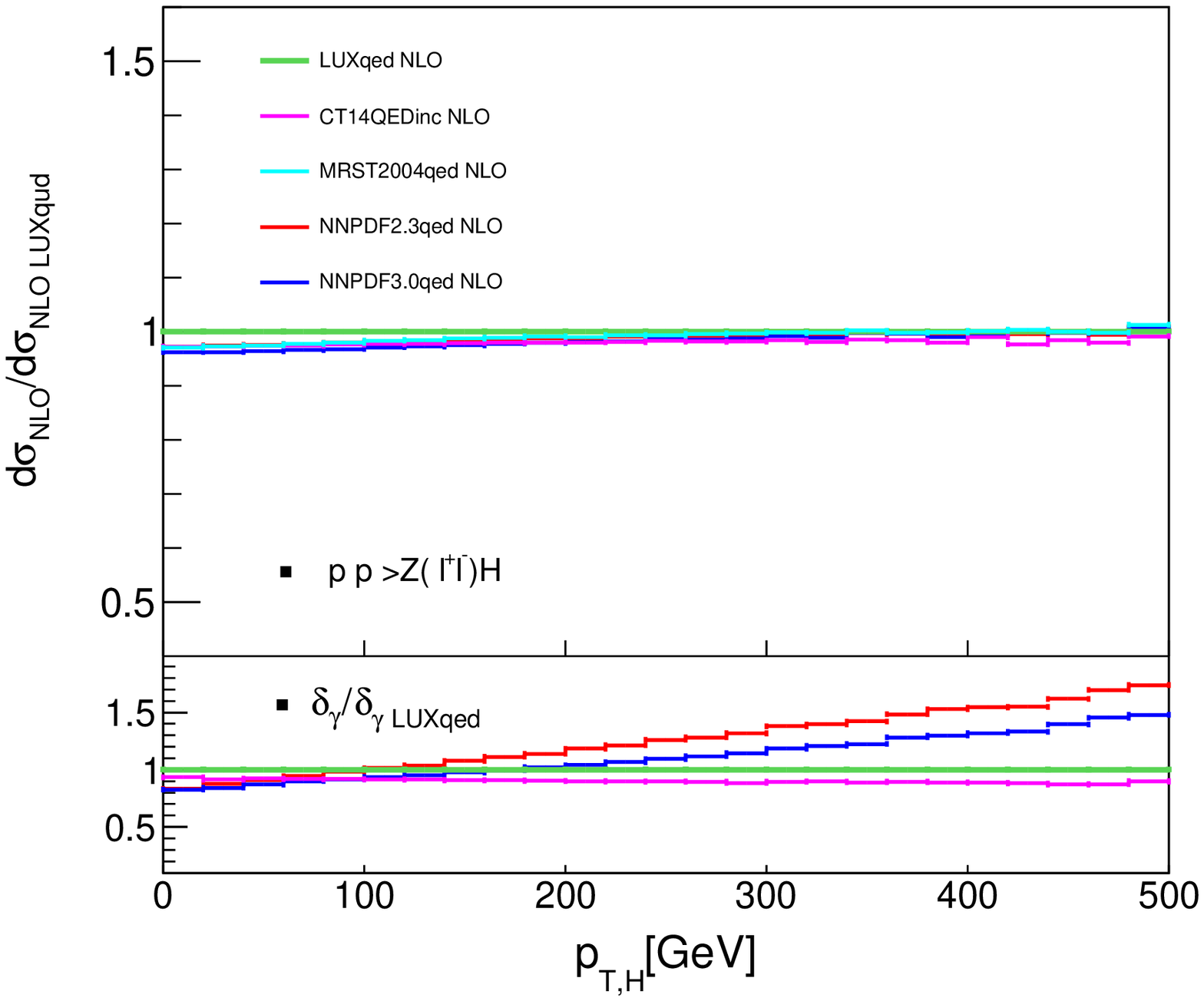}
\end{minipage}
}
\caption{
Differential cross section distributions (top left  panel) both at LO (lower curves) and NLO (upper curves) and ratio of $ d\sigma_{NLO}/d\sigma_{NLO\;LUXqed}$ (top right  panel)  of the Higgs boson in $Z(l^+l^-)H$ production after applying the selection cuts in section \ref{cuts} at the LHC at 13 TeV for the various PDFs.
 The bottom left  panel shows the electroweak $\delta_{EW}$ and photon corrections $\delta_\gamma =d\sigma_\gamma/d\sigma_{LO}$. The bottom right  panel is the ratio of $\delta_\gamma/\delta_{\gamma LUXqed}$.
}
\label{fig-4-44}
\end{figure}

In this subsection we consider the differential cross sections after applying cuts to the charged leptons.
We plot the transverse-momentum distributions for Higgs production in $pp \to W^\pm /Z + H  \to l\nu_l \//l^+l^- +H$ at LO and NLO in QCD for Higgs boson mass $M_H=125$ GeV with the kinematical cuts and set up described in Section \ref{cuts}.
The $p_T(H)$ distributions of the Higgs boson are shown in Figs. \ref{fig-2-22}, \ref{fig-3-33} and \ref{fig-4-44} for various PDFs.
From the NLO (upper) and LO (lower) $p_T(H)$  distribution  curves  we see that
the NLO correction is small at small $p_T(H)$, and increases slightly with $p_T(H)$  as the cross section falls rapidly.
 The upper left  plot is the  transverse-momentum distributions of Higgs boson  production at LO (lower curves) and NLO (upper curves) including NLO QCD and EW corrections. The lower left plot is the  photon correction $\delta_\gamma =d\sigma_\gamma/d\sigma_{LO}$ and EW corrections $\delta_{EW}$  for various PDFs.
The EW correction for the process $pp \to H+ W(\to l\nu_l)$ grows further negative to (10--20)\% as $p_{T}(H)$ increases. The photon-induced corrections have a tendency to grow as well, but can reach only 2.5\% level.
In the top right plot we show transverse-momentum distributions of various PDF predictions to LUXqed prediction. The bottom right panel is the ratio  of $\delta_\gamma/\delta_{\gamma LUXqed}$.

\section{Conclusions}\label{conclusion}

We have calculated the NLO QCD and EW corrections to
Higgs boson production in association with $W$ and $Z$~bosons at the LHC at 13 TeV and with Higgs boson mass $M_H=125$ GeV for various PDFs using the HAWK Monte Carlo tool within the $G_\mu$-scheme.
The total cross sections and fiducial cross sections for the processes $pp \to W^\pm  + H  \to l\nu_l \ +H$ and $pp \to Z + H  \to l^+l^- +H$  were given in Tables
 \ref{ta:xsection_nocuts}, \ref{ta:xsection_nocuts2}, \ref{ta:xsection_nocuts3}, \ref{ta:xsection_withcuts4}, \ref{ta:xsection_withcuts5}, and \ref{ta:xsection_withcuts6}.
The EW corrections to  $W^+(\to l^+\nu_l)H$, $W^- (\to l^- \bar \nu_l)H$ and $Z(\to l^+l^-)H$ productions decrease the total cross sections by about -7.4\%, -7.3\% and -5.1\% respectively.
The NLO QCD corrections to  $W^+(\to l^+\nu_l)H$, $W^- (\to l^- \bar \nu_l)H$ and $Z(\to l^+l^-)H$ productions increase the total cross sections by about 17.5\%, 17.0\% and 17.0\% respectively.
Our photon-induced process fiducial cross sections in Tables \ref{ta:xsection_withcuts4}, \ref{ta:xsection_withcuts5}, and \ref{ta:xsection_withcuts6} agree well with the results of Ref.~\cite{2016-008cc} in their Tables 5.7, 5.8, and 5.9; and our EW corrections in Tables  \ref{ta:xsection_nocuts} - \ref{ta:xsection_nocuts3}, and \ref{ta:xsection_withcuts4} -  \ref{ta:xsection_withcuts6} agree well with the results of Ref.~\cite{2016-008cc} in their Tables 5.3-5.5 and 5.7-5.9.
The PDF uncertainties of the  photon-induced process cross sections are reduced significantly
for NNPDF2.3qed, NNPDF3.0qed,  MRS2004qed, and CT14QEDinc photon PDF sets after including the  CMS  8~TeV data.
Especially, the CMS data can have a very large impact in reducing the  NNPDF photon PDF
errors.

\section{Acknowledgments}
We thank Carl Schmidt and C.-P. Yuan for many very helpful discussions.
This work is supported by the National Natural Science Foundation of China under the Grant No. 11465018.

\clearpage

\end{CJK*}
\end{document}